\documentclass[twocolumn,showpacs,preprintnumbers,amsmath,amssymb,prl]{revtex4}

\usepackage{graphicx}
\usepackage{dcolumn}
\usepackage{bm}

\begin{document}

\preprint{}

\title{Geometrical Resonance Conditions for THz Radiation from \\
the Intrinsic Josephson Junctions in \bm{\mathrm{Bi$_2$Sr$_2$CaCu$_2$O$_{8+\delta}$}}}

\author{Manabu Tsujimoto}%

\author{Kazuhiro Yamaki}%

\author{Kota Deguchi}%

\author{Takashi Yamamoto}%

\author{Takanari Kashiwagi}%

\author{Hidetoshi Minami}%

\author{Masashi Tachiki}%

\author{Kazuo Kadowaki}

\affiliation{%
Institute of Materials Science, Graduate School of Pure \& Applied Sciences, University of Tsukuba,
1-1-1, Tennodai, Tsukuba, Ibaraki 305-8573, Japan
}%

\author{Richard A. Klemm}

\affiliation{Department of Physics, University of Central Florida, Orlando, Florida 32816, USA}

\date{\today}

\begin{abstract}
Subterahertz radiation emitted from a variety of short rectangular-, square-, and disk-shaped mesas of intrinsic Josephson junctions fabricated from a Bi$_2$Sr$_2$CaCu$_2$O$_{8+\delta }$ single crystal was studied from the observed $I$-$V$ characteristics,  far-infrared spectra, and  spatial radiation patterns.  In all cases, the radiation frequency satisfies the conditions both for the  \textit{ac} Josephson effect and for a mesa cavity resonance mode.  The integer higher harmonics observed in all spectra imply that the \textit{ac} Josephson effect plays the dominant role in the radiation.
\end{abstract}

\maketitle


After the first report of  intense continuous terahertz (THz, $1 \text{THz} = 10^{12}\ \text{Hz} $) electromagnetic (EM) waves emitted from the intrinsic Josephson junctions (IJJs) in the high temperature superconductor (HTSC) Bi$_2$Sr$_2$CaCu$_2$O$_{8+\delta }$ (Bi-2212)  by Ozyuzer \textit{et~al.}~\cite{Ozyuzer07} with remarkably higher intensity than previously generated from Josephson junctions~\cite{Langenberg65}, a great deal of interest has been drawn not only to the physical mechanism of the radiation but also to the possible variety of applications in the vast fields of science and technology.  Nondestructive inspections, medical diagnostics, high speed communications, imaging technologies for security and defense, \textit{etc.}, are  potential candidates among them.  Presently available THz sources such as those using parametric generation or pulsed-current methods based on  semiconductor and/or laser technology are rather weak in output power, and are mostly pulsed with incoherent waves.  Although it is possible to generate output  power of even $\sim \mu $W in a two-dimensional array of  Josephson junctions~\cite{Barbara99},  frequencies above a few hundred GHz cannot be achieved with  conventional superconductors due to the small energy gap of a few meV (1~meV $\Leftrightarrow$ 483.5979~GHz).  However, single crystalline  Bi-2212 has an approximately 10 times larger energy gap, so that in principle it can  be employed to reach frequencies of several THz.  Furthermore, it was shown~\cite{Ozyuzer07} that an entirely new  mechanism for THz EM wave generation occurs in Bi-2212 due to its highly anisotropic layered structure of  IJJs~\cite{Kleiner92}.

\begin{figure}[t] 
	\includegraphics[bb=0 0 685 540, width=0.40\textwidth ,clip]{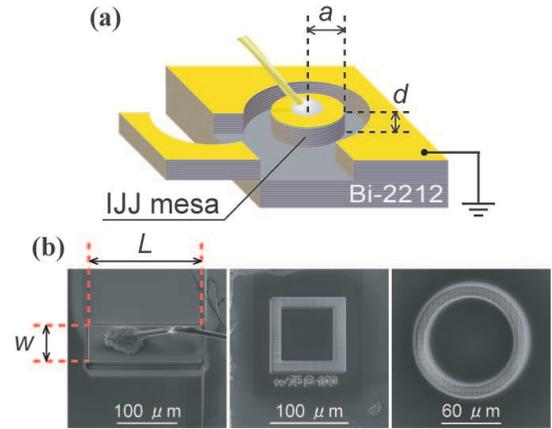} \\
	\caption{(a) Sketch of a cylindrical disk IJJ mesa.  The grazing angle forming the groove of 10~$\mu $m width is approximately $5^{\circ}$.  (b) Scanning ion microscope images of a rectangular (left), square (center), and disk (right) mesa.
}
	\label{fig:1}
\end{figure}%

\begin{table*}
	\caption{
\label{tab:1} List of the parameters of the IJJ mesa samples.  See text.
}

\begin{ruledtabular}
\begin{tabular}{llcccccc}
No. & Geometry & $a$ [$\mu $m] & $w $ $(L) $ [$\mu $m] & $d$ [$\mu $m] & $T_c $ $(\Delta T_c)$ [K] & $\delta T$ [K] & $j_c $ [A/cm$^2$] \\ \hline

D1 & Disk & 33.9--38.9 & & 1.4 & 88.3 (1.9) & 20--50 & 200 \\
D2 & Disk & 48.9--51.5 & & 1.5 & 88.5 (2.0) & 35--45 & 55 \\
D3 & Disk & 61.5--65.0 & & 1.6 & 85.9 (4.0) & 10--50 & 190 \\
S1 & Square & & 66.7--75.7, 70.9--81.6 & 1.7 & 87.7 (1.1) & 30--40 & 180 \\
R1 & Rectangle & & 59.7--64.4 (200) & 1.5 & 82.3 (2.5) & 10--25 & 170 \\

\end{tabular}
\end{ruledtabular}
\end{table*}

Here, we present direct unambiguous evidence that the \textit{ac} Josephson effect is the  driving mechanism for the radiation, in which the fundamental frequency $f_1$ equals the frequency $f_J=2eV/(Nh)$ of the \textit{ac} Josephson effect, where $V$ is the applied \textit{dc} voltage across the $N$ IJJs, $e$ is the electric charge, and $h$ is Planck's constant. $f_1$ is also in resonance with a frequency $f_{mp}$ of a thin EM cavity mode, which is inversely proportional to the minimal mesa cross-section dimension.  We demonstrate THz (or sub-THz) radiation from mesas of various geometrical shapes, using rectangular-, square-, and cylindrical disk-shaped mesas fabricated by focused ion beam (FIB) milling.  We first examine the observed radiation frequencies as  functions of the geometrical sizes of the mesas, in order to analyze the data by the above two radiation conditions~\cite{Ozyuzer07,Kadowaki07}.  The cylindrical geometry is particularly interesting for understanding the radiation, because the standing EM cavity  mode frequencies $f_{mp}^{\text{c}}$ are determined by the zeroes of the first derivatives of standard Bessel functions, which are incommensurate with the  harmonic \textit{ac} Josephson frequency spectrum $nf_J$ for integer $n$\cite{Klemm09}.  Hence, only one frequency can equal both a cylindrical cavity mode frequency and an \textit{ac} Josephson frequency.  In contrast, for rectangular mesas, each member  of a subset  of higher cavity modes of frequencies $f^{\text{r}}_{mp}$ can be degenerate with an harmonic of the fundamental \textit{ac} Josephson frequency $f_J$~\cite{Klemm09}.  For the rectangular mesas used in all previous studies~\cite{Ozyuzer07,Kadowaki07,Minami09}, it was therefore not possible to unambiguously determine the primary mechanism of the radiation.  Comparisons of the angular distribution of the observed radiation with that predicted can also provide information as to the relative importance of the two mechanisms for the radiation~\cite{Hu08,Klemm09,Kadowaki10}.

\begin{figure}[t] 
	\centering
	\includegraphics[bb=0 0 670 320, width=0.46\textwidth ,clip]{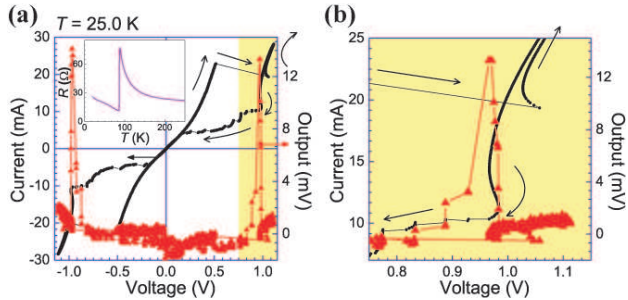} \\
	\caption{(a) The $I$-$V$ characteristics (left scale) and  output radiation intensity  detected by the Si-bolometer (right scale) from  disk mesa D3 at the bath temperature $T=25.0$~K of its maximal radiation intensity are shown. The inset shows the $c$-axis $R$-$T$ curve.  (b) Details of the shaded high bias region in Fig.~\ref{fig:2}(a) where the emission is observed.
}
	\label{fig:2}
\end{figure}%

The single crystals of Bi-2212 used in the present studies were prepared by a floating zone technique~\cite{Mochiku94}.  The as-grown single crystals were annealed at 650~$^{\circ} $C for 24 h in argon gas mixed with 0.1\% oxygen in order to obtain  underdoped crystals.  The temperature $T$ dependence of the $c$-axis resistance ($R$-$T$ curve) shows the behavior typical  of  slightly underdoped Bi-2212 crystals with a critical temperature $T_c\approx87$~K.  A small piece of a cleaved crystal  $\sim $50~$\mu $m  thick was glued onto a sapphire substrate by silver paste.  Next, silver and gold thin layers were evaporated onto the surface.  Then, a groove of  width 10~$\mu $m and  depth of $\sim $2~$\mu $m was patterned by  FIB milling,  making an islandlike terrace as sketched in Fig.~\ref{fig:1}(a).  At the end of the process, a 10~$\mu $m gold wire  electrode was connected to the center of the top metal layer of the mesa by silver paste.  The sample dimensions and profile curves were measured by an atomic force microscope  and the results are presented in Table~\ref{tab:1}.  The cross-sectional profiles are considerably slanted and rounded at the edges, resulting in  approximately trapezoidal shapes (not shown here)\cite{Kadowaki07,Kadowaki10}.  Typically the top cross-section is approximately 10--20\% smaller than the bottom one.  Thin mesas with three different cross-sectional shapes were studied: one rectangular, one square, and three circular disks with properties listed in Table~\ref{tab:1}.  The total number $N \sim  1000 $ of IJJs is roughly estimated from each mesa height $d$.

 The  $I$-$V$ characteristics of  disk mesa D3, measured by sweeping the bias \textit{dc} current $I$ along the $c$-axis at the bath temperature $T=25.0$~K, are displayed in Fig.~\ref{fig:2}(a) together with the radiation intensity detected by the Si-bolometer.  In Fig.~\ref{fig:2}(b) the radiation region is shown in detail in an expanded scale.  All mesas that emit THz radiation  studied to date have a critical current density $j_{\text{c}} $ ranging between 55 and 200~A/cm$^{2}$ as listed in Table~\ref{tab:1}.  The strongest radiation occurs on the return branch of the outermost $I$-$V$ curve.  In most cases, the THz radiation was observed in a very narrow voltage range.  For example,  emission occurs for sample D3 between 0.96 and 0.98~V, corresponding to $I\approx 10.8$~mA.  Then, the radiation suddenly stops due to a jump to another $I$-$V$ characteristic  branch.  The radiation power density  at the detector just before jumping is  estimated to be 1.3~nW/cm$^2$, the same order of magnitude as obtained previously from  rectangular mesas~\cite{Kadowaki07}.  If the interbranch  jump did not occur, the radiation intensity would likely have grown much stronger~\cite{Kadowaki10}.

It is also interesting to note that the emission usually occurs within a sample-dependent range $\delta T$ of the base temperature of the mesa,  as summarized in Table~\ref{tab:1}.  For example,  mesa D3  emits between 10 and 50~K.  Since  the constant $I\approx 11$~mA is fed into  mesa D3, it is inevitably heated at a rate of about 11~mW, corresponding to the enormous heating power density of 8.3~kW/cm$^3$.  This huge heating power density cannot be removed quickly enough from the mesa to maintain equilibrium, resulting in a considerable rise of the mesa temperature.  This local heating may induce a chaotic nonequilibrium state and may adversely affect the THz radiation~\cite{Wang09,Kurter09}.  Since the heat conduction is progressively worse  as $T$ is lowered~\cite{Yasuda97,Fenton02} and the gap vanishes as $T\rightarrow T_c$, these features may account for the peculiar temperature dependence of the radiation intensity.

\begin{figure}[t] 
	\centering
	\includegraphics[bb=0 0 435 540, width=0.40\textwidth ,clip]{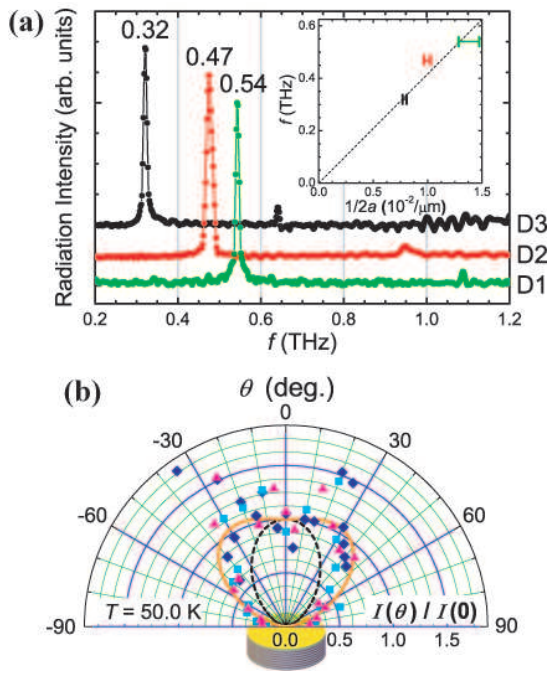} \\
	\caption{(a) The radiation spectra measured by the FTIR spectrometer for three disk mesas.  The inset shows the observed frequencies versus $1/(2a)$.  The error bars reflect  the trapezoidal cross-section profile.  The dashed line represents the resonance frequency expected from the cylindrical cavity TM (1,~1) mode with $\epsilon = 17.6$.  (b) Polar plot of the radiation intensity for disk mesa  
D3.  Different symbols correspond to different runs.  The  solid  
orange and dashed black curves are the best fit to the dual-source  
model and its cavity component, respectively.
}
	\label{fig:3}
\end{figure}

In Fig.~\ref{fig:3}(a), the far-infrared spectra of the THz radiation from the disk mesas, as measured by a Fourier transform infrared (FTIR) spectrometer, are shown.  In the inset to Fig.~\ref{fig:3}(a), the fundamental radiation frequency $f_1=f_J$ is plotted as a function of $1/(2a) $, where $a $ is the mesa radius.  The dashed line  represents the resonance frequency $f_{J} =f^{\text{c}}_{11}=\chi_{11}c_0/(2\pi\sqrt{\epsilon} a)$,  where $\chi_{11} =1.841$ for the TM (1,~1) mode, $\epsilon$ is the dielectric constant of Bi-2212, and $c_0$ is the speed of light in vacuum, predicted by the cavity resonance model~\cite{Derneryd79}, assuming the cylindrical cavity boundary condition  $H_{\phi}|_{\rho=a}=(\partial E_{z}/\partial \rho)|_{\rho=a}=0$. The radiation frequency $f_1$ is clearly proportional to $1/(2a)$. The data were fitted with  $\epsilon = 17.6$,  in good agreement with previous results~\cite{Kadowaki07,Kadowaki10}.  Note that this $\epsilon $ value is about 50\% larger than that obtained from infrared spectroscopy~\cite{Tajima93}.  In addition, for each disk mesa, the second harmonic at $f_2=2f_J$ is clearly visible, and these frequencies are easily distinguishable from those of the nearest higher disk cavity modes~\cite{Klemm09}, providing unambiguous experimental evidence that the uniform \textit{ac} Josephson current is the primary radiation source.

In order to further understand the excitation mode inside the disk mesa, the  radiation  intensity ${\cal I}$ was measured at various detection angles $\theta $, relative to the normal to the mesa.  In Fig.~\ref{fig:3}(b),  ${\cal I}(\theta)$ for  disk mesa D3 is presented.  The shadowing effect  of the radiation from the superconducting Bi-2212 crystal wall outside the groove is expected to be negligibly small to  first approximation.  Therefore, the following characteristic features are noted: First,  $\mathcal{I}(\theta) $ is strongly anisotropic, having a maximum around $\theta=\theta_{\text{max}} = 20\text{--}35^{\circ}$ from the top ($\theta=0^{\circ}$), where a local minimum occurs with  intensity ratio $\mathcal{I}(\theta=0^{\circ})/\mathcal{I}_{\text{max}}(\theta = 20^{\circ})=0.50\text{--}0.65$.  This shallow minimum feature is almost the same as for  rectangular mesas, although  $\theta_{\text{max}}$ is somewhat less than the corresponding rectangular mesa value~\cite{Kadowaki10}.  Second, ${\cal I}(\theta)$ rapidly diminishes as $\theta$ approaches 90$^{\circ}$.  ${\cal I}_{\text{cav}}(\theta)$ calculated~\cite{Klemm09} by assuming radiation from the TM (1,~1) cavity resonance mode alone is shown  by the dashed black curve in Fig.~\ref{fig:3}(b).  Clearly,  the calculated ${\cal I}_{\text{cav}}(\theta)$ does not fit  the experimental data, especially near to $\theta=0^{\circ}$, where ${\cal I}_{\text{cav}}(\theta)$ is a maximum~\cite{Klemm09}.  This disagreement can be reduced by introducing a superposition of the radiation from the uniform \textit{ac} Josephson current source with the same Josephson frequency, as described for rectangular mesas~\cite{Klemm09,Kadowaki10}.  The experimental data are  better fitted by the dual-source model ${\cal I}_{\text{dual}}(\theta)$ with mixing parameter $\alpha=1.44$~\cite{Klemm09}, corresponding to 58\% of the radiation arising from the uniform \textit{ac} Josephson current source, as shown by the solid orange curve in Fig.~\ref{fig:3}(b).

It is significant that the intensity from the uniform source is comparable to that of the fundamental cavity mode source, as in rectangular mesas~\cite{Kadowaki10}.  Since the fundamental cavity mode radiation is enhanced by the cavity quality Q-value, a similar enhancement must occur for the uniform source radiation\cite{Tachiki09}.  This suggests that the radiation from the uniform \textit{ac} Josephson current in the $N$ junctions is coherent, amplifying the output by a factor of order $N^2$\cite{Barbara99}.  This interpretation is strongly supported by the observation of only integral higher harmonics of the fundamental frequencies  shown in Fig.~\ref{fig:3}(a).  Neither higher cylindrical cavity excitation frequencies of the Bessel type nor subharmonics were observed.  The  
higher harmonics are naturally present in the uniform Josephson  
current, which radiates coherently, but the nonuniform part of which  
only can excite one cylindrical cavity mode.  This experimental evidence clarifies unambiguously that the THz radiation is mainly generated by the uniform mode of the \textit{ac} Josephson current, in sharp contrast to a number of theoretical predictions~\cite{Koshelev08-1,Lin08-1,Lin08-2,Hu08}.  Note that although the harmonics arise solely from the uniform \textit{ac} Josephson current source, the larger intensity fundamental frequency is a mixture of both the uniform \textit{ac} Josephson and nonuniform cavity sources.  Further studies of the polarization, coherence, and spatial radiation patterns of the higher harmonics could provide  supporting  information for these conclusions.

\begin{figure}[htpb] 
	\centering
	\includegraphics[bb=0 0 500 355, width=0.33\textwidth ,clip]{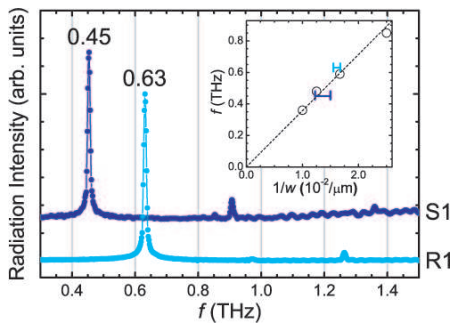} \\
	\caption{The radiation spectra for the square (S1) and rectangular (R1) mesas.  The inset shows the observed frequencies versus the narrower width inverse $1/w$, where the circles are previous data~\cite{Kadowaki10}.  The black dashed  line represents the resonance frequency expected for the rectangular cavity TM (0,~1) mode with $\epsilon = 17.6$.
}
	\label{fig:4}
\end{figure}

The resonance frequencies of the TM ($m,p$) modes of a rectangular cavity of width $w$ and length $L$ are $f_J=f^{\text{r}}_{mp} = (c_0/2\sqrt{\epsilon})\sqrt{(m/L)^2+(p/w)^2} $~\cite{Kadowaki10}, whereas for cylindrical cavities, they are $f_J=f^{\text{c}}_{mp}=\chi_{mp}c_0/(2\pi \sqrt{\epsilon}a)$, where $\chi_{mp}$ is the $p$th zero of the derivative of $J_m(z)$, the standard Bessel function of order $m$~\cite{Klemm09}.  Theoretically, the fundamental modes are the TM (1,~0) and TM (1,~1) modes for  rectangular and cylindrical cavities, respectively.  Although this is consistent with experiment for cylindrical mesas, it is very curious that the TM (1,~0) mode for  rectangular mesas was never observed in  experiment~\cite{Ozyuzer07,Kadowaki07,Kadowaki10}.  Since the mesa may be considerably heated  by the \textit{dc} current, it is very likely that the inhomogeneous heat distribution inside it prevents standing EM wave formation, particularly along the longer rectangular dimension.  However, such hot spots with $T$ even exceeding $T_c$ observed by Wang \textit{et~al.}~\cite{Wang09} do not seem to be a problem in our experiments, since the fundamental frequencies of our disk and square mesas excellently obey the linear relation of the cavity resonance frequencies with  $1/a$ and $1/w$, respectively.  The discrepancies in relatively long rectangular mesas with $w \ll L = $ 300--400~$\mu $m may have a different origin.  In order to check this, we fabricated  rectangular mesas with smaller  $L/w$ ratios, as shown in the left and center scanning ion microscope images in Fig.~\ref{fig:1}(b).  The spectroscopic emission data from these mesas  are presented in Fig.~\ref{fig:4}.  It is obvious from the inset to Fig.~\ref{fig:4} that the frequency at 0.63 THz of the rectangular mesa R1 with  $L/w\approx 3.1\text{--}3.3 $  obeys the TM (0,~1) cavity resonance mode very well, but definitely not the TM (1,~0) resonance mode.  These rather surprising observations strongly suggest that the formation of EM standing waves are restricted  by  other as yet undetermined reasons, and may not be excited below some cutoff frequency.  We propose that this cutoff frequency may be the Josephson plasma frequency $f_{p} = c_{0} /(2\pi\sqrt{\epsilon } \lambda _{c})$, where  $\lambda _{c}$ is the  $c$-axis superconducting penetration depth.

In summary, we  studied the THz radiation generated from  IJJ mesas of Bi-2212 with different geometrical shapes.  Our experimental results clearly demonstrate the validity  of the cavity resonance model for the fundamental frequency mode of thin square and cylindrical mesas.  The spatial radiation pattern cannot be explained by the cavity resonance model alone, but requires a substantial contribution from the uniform \textit{ac} Josephson current source.  More importantly, the frequency spectra obtained exhibit higher integral harmonics of the fundamental $f_1$, which cannot be obtained from higher cavity resonance modes in cylindrical cavities, providing unambiguous experimental evidence that the uniform \textit{ac} Josephson current is the primary radiation source.  Although heating effects may  significantly alter the $I$-$V$ characteristics, they do not greatly affect the two necessary radiation conditions:  the \textit{ac} Josephson relation, $f_{1}=f_{J} = 2eV/Nh$, and  the geometrical cavity resonance condition $f_1=f_{11}^{\text{c}}$ or $f_1=f^{\text{r}}_{01}$, for cylindrical or rectangular cavities, respectively, each inversely proportional to the minimal cross-sectional dimension.  We propose a further radiation condition that $f_1 > f_p$, the Josephson plasma frequency.

The authors deeply thank X. Hu, S. Lin, A. Koshelev, M. Matsumoto, T. Koyama, M. Machida, S. Fukuya, and K. Ivanovic for valuable discussions and Y. Ootuka, A. Kanda, I. Kakeya, H. Yamaguchi, N. Orita, T. Koike, and B. Markovic for their technical assistance.
This work has been supported in part by CREST-JST (Japan Science and Technology Agency), WPI (World Premier International Research Center Initiative)-MANA (Materials Nanoarchitectonics) project (NIMS) and Strategic Initiative category (A) at the University of Tsukuba.




\end{document}